\documentstyle[aps,eqsecnum]{revtex}
\newcommand{\lsim}{\raise.35ex\hbox{$<$}\kern-0.75em\lower.5ex\hbox{$\sim$}}
\newcommand{\gsim}{\raise.35ex\hbox{$>$}\kern-0.75em\lower.5ex\hbox{$\sim$}}

\begin{document} \draft 
\title{Spin-gap phase in nearly-half-filled one-dimensional conductors
coupled with phonons}
\author{K. Yonemitsu\cite{present}} 
\address{Department of Applied Physics, 
Tohoku University, Sendai 980, Japan}
\author{M. Imada}
\address{Institute for Solid State Physics,
University of Tokyo, Tokyo 106, Japan}
\date{\today}
\maketitle

\begin{abstract}
Asymptotic properties of nearly-half-filled one-dimensional conductors 
coupled with phonons are studied through a renormalization group method.
Due to spin-charge coupling via electron-phonon interaction, the spin 
correlation varies with filling as well as the charge correlation.
Depending on the relation between cut-off energy scales of the Umklapp 
process and of the electron-phonon interaction, various phases appear.
We found a metallic  
phase with a spin gap and a dominant charge-density-wave correlation 
near half filling between 
a gapless density-wave phase (like in the doped repulsive Hubbard model) 
and a superconductor phase with a spin gap.
The spin gap is produced by phonon-assisted backward scatterings 
which are interfered with the Umklapp process constructively or 
destructively depending on the character of electron-phonon coupling.
\end{abstract} 
\pacs{72.15.Nj,71.38.+i,71.45.Lr,71.30.+h}

\section{INTRODUCTION}

Metallic states of doped copper oxides and related materials  
are extensively studied to clarify the characteristics of strongly correlated 
electron systems.
The ``spin gap'' feature of the underdoped compounds\cite{magnetic} is one of 
the most peculiar features of them, but its origin is not yet clear.
In a model with strong on-site attraction, there would be a temperature range 
in which local singlet pairs are formed with finite binding energy but the 
system is still not superconducting.  
In the case of the strong on-site repulsion, however, the spin gap formation 
must be the consequence of a subtle balance. 
As a fundamental property of strongly correlated 
materials, it is worth investigating the spin excitation gap in metals 
from various points of view.
It is pointed out that a spin gap may play an important role both in 
determining the type of a metal-insulator transition\cite{Imada1} and 
in the mechanism of the high-$T_{\rm c}$ superconductivity.\cite{Imada2}

It has been pointed out that electron-phonon interaction can drastically 
change {\em local\/} quasiparticle characters of nearly-half-filled electron 
systems {\em when the on-site repulsion is strong\/}.\cite{YBL,ZS,Fehske}
In this paper, we investigate asymptotic behaviors of correlation functions 
by integrating out short-distance degrees of freedom in space and time.
Repulsive interaction and retarded attraction mediated by phonons are 
treated on an equal footing.
Our studies are limited here to one dimension where a bosonization 
technique\cite{Solyom,Emery} can be used to derive renormalization group 
equations and to take the filling factor explicitly into account.
Attention is paid to tendency for the formation of a spin gap 
near half filling 
and connections to the insulator phase at half filling 
and to a possible superconductor phase away from half filling.

Without electron-phonon interaction, spin and charge excitations are separated 
through bosonization which only describes low-energy excitations.
Phonons couple with both spin and charge so that they are interfered with 
each other until electron-phonon interaction is integrated out 
at energy of the order of phonon frequency.
This spin-charge coupling at finite energy leads the spin and 
charge correlations to depend on each other.
Technically, it leads the equivalent two-dimensional classical system 
to have interacting Burgers vectors.\cite{Voit}

At half filling, the Umklapp process tends to open a charge gap.
Away from half filling, as the energy scale is made smaller, 
the Umklapp process 
becomes ineffective at finite energy which 
vanishes in the limit of half filling.
Therefore the charge excitation is always gapless except half filling.
Nevertheless the Umklapp process at finite energy modifies the 
charge correlation and also the spin correlation 
through the electron-phonon coupling.
Thus the spin excitation is controlled by the relation between 
the cut-off energy scale of the Umklapp process and that of electron-phonon 
interaction.
In terms of two-dimensional classical systems, the interactions between 
Burgers vectors originating from these two processes have different shapes 
and different length scales so that they are cut off in different manners.
We numerically solve renormalization group equations for the correlation 
exponents and scattering parameters (fugacities in the classical system)
and find a new phase which is made possible by both of these two processes.

In this paper, we study a bosonized Hamiltonian in which 
the standard electronic part is coupled with dispersionless phonons 
so that back and forward scatterings occur exchanging phonons 
as well as directly between electrons.
In the electronic part, we consider on-site repulsion.
The scattering parameters of electron-electron interaction are 
taken from the lowest-order perturbation with respect to on-site 
repulsion. 
This is because the exactly known, fixed-point values of the pure 
Hubbard model\cite{Schulz} are no longer valid due to modified scaling 
properties.
For a phonon-assisted backward scattering, we take 
a continuum version of the on-site Holstein coupling\cite{Holstein} or 
that of the intersite Su-Schrieffer-Heeger (SSH) coupling.\cite{SSH}
Furthermore, we add a phonon-assisted forward scattering 
which appears in the continuum version of the Holstein model 
in order to make a singlet superconductor phase possible.
It is numerically confirmed that this addition to the SSH case does not affect 
the spin correlation so much.

The renormalization group equations we obtain deal with the 
lowest-order perturbations, so that the fixed point of  
either correlation exponent at zero should correspond to the Luther-Emery 
state\cite{Luther} with a gap in the corresponding channel.
At half filling, our results for the critical coupling strengths 
for the Holstein model and the SSH model 
(without a phonon-assisted forward scattering) to have 
a charge density wave with both spin and charge gaps 
are consistent with the critical coupling strengths 
for dimerization obtained analytically and by the quantum Monte Carlo 
method.\cite{Hirsch} 
Away from half filling, the lattice is not statically distorted 
for purely one-dimensional phonons with finite frequency.
More realistically, three-dimensionality of the crystal lattice 
would effectively freeze a periodic lattice distortion, if any.
In this paper, we do not consider such a case but study effects 
of {\em dynamical\/} phonons on the spin excitation gap in correlated 
metals.

Far away from half filling where the Umklapp process is ineffective, 
a phase diagram is divided basically into 
a gapless phase with a dominant density-wave correlation 
as in the doped repulsive Hubbard model 
and a singlet superconductor phase with a spin gap for large enough 
electron-phonon interaction.
Near half filling, we found a phase with a spin gap and a dominant 
charge-density-wave correlation between the two phases and call it a 
spin-gap charge-density-wave (SG-CDW) phase.
This phase does not have a charge gap and is metallic with a finite 
Drude weight.
The wave number of the SG-CDW is incommensurate to the lattice 
so that it should be metallic due to a phason mode.\cite{LeeRiceAnderson}
The appearance of the SG-CDW phase is thus reasonable and related to the 
similar phase in Ref.~\cite{Voit} where an Umklapp process and a 
phonon-assisted forward scattering are not considered.
In this paper, we clarify its relation with the Mott insulator phase 
and with a superconductor phase 
by treating both electron-electron and electron-phonon 
interactions at and away from half filling.
Furthermore, effects of the combination of these two kinds of interactions 
near half filling are demonstrated to depend strongly on the form of 
electron-phonon interaction.
It is reminiscent of the two-dimensional 
case\cite{YBL} in that a system with strong on-site repulsion is more 
sensitive to an intersite electron-phonon coupling than an on-site one 
for a small phonon frequency.

This paper is organized as follows: 
Sec.~\ref{sec:model} introduces the bosonized model.
Relations of electron-phonon parameters to the 
$g$-ology parameters are examined.
Sec.~\ref{sec:renormalization} outlines the derivation of 
renormalization group equations.
Some modifications to previous works are described.
It is confirmed that the antiadiabatic limit is correctly reproduced.
Sec.~\ref{sec:results} shows phase diagrams for different 
electron-phonon couplings and phonon frequencies.  
The evolution of parameters with changing scales is demonstrated for 
typical examples to clarify the roles played by different processes.
Sec.~\ref{sec:summary} summarizes the present work.
Part of the results presented in this paper were reported briefly 
elsewhere.\cite{YI}

\section{MODEL}\label{sec:model}

We consider a Tomonaga-Luttinger model extended to include the
spin and charge degrees of freedom, a backward scattering, 
an Umklapp process, and phonon-assisted back and forward scatterings.
The electronic part is a standard one.\cite{Solyom,Emery}
Assuming weak interactions, one can linearize the electronic dispersion 
relation around the Fermi points.  
The kinetic part is written as a term bilinear in boson operators 
for spin and charge densities,
\begin{equation}
\sigma_j(p)=\frac1{\sqrt{2}}\left( 
\rho_{j\uparrow}(p) - \rho_{j\downarrow}(p) \right)
\;,
\end{equation}
\begin{equation}
\rho_j(p)=\frac1{\sqrt{2}}\left( 
\rho_{j\uparrow}(p) + \rho_{j\downarrow}(p) \right)
\;,
\end{equation}
respectively, with the Fourier components of the electron density 
operators,
\begin{equation}
\rho_{j,s}(p)=\sum_k c_{j,k+p,s}^\dagger c_{j,k,s}
\;,
\end{equation}
for right ($j$=1) and left ($j$=2) movers with spin $s$.
They obey commutation relations,
\begin{equation}
\left[ \nu_1(-p), \nu_1(p') \right] = 
\left[ \nu_2(p), \nu_2(-p') \right] = \frac{pL}{2\pi} \delta_{p,p'}
\;,
\ \ \ \ \ \ (\nu = \sigma, \rho)
\:.
\end{equation}
A backward scattering between parallel spins ($g_{1\parallel}$),  
a forward scattering which couples the two branches, $j=1,2$, 
($g_2 \equiv g_{2\parallel}=g_{2\perp}$), and 
a forward scattering within a single branch ($g_4 \equiv g_{4\perp}$, 
$g_{4\parallel}$ is set at zero.) are also 
written as bilinear terms.
By combining them with the kinetic part, 
the spin and charge parts of the bilinear terms read
\begin{equation}
H_{\sigma,0} = \frac{2\pi v_{\rm F}-g_4}{L}\sum_{p>0} 
	\left[ \sigma_1(p) \sigma_1(-p) + \sigma_2(-p) \sigma_2(p) \right] \\ 
	- \frac{g_{1\parallel}}{L}\sum_{p>0} 
	\left[ \sigma_1(p) \sigma_2(-p) + \sigma_1(-p) \sigma_2(p) \right]
\:,\label{eq:sigma_0}
\end{equation}
\begin{equation}
H_{\rho,0} = \frac{2\pi v_{\rm F}+g_4}{L}\sum_{p>0} 
	\left[ \rho_1(p) \rho_1(-p) + \rho_2(-p) \rho_2(p) \right] \\ 
	- \frac{g_{1\parallel}-2g_2}{L}\sum_{p>0} 
	\left[ \rho_1(p) \rho_2(-p) + \rho_1(-p) \rho_2(p) \right]
\:,\label{eq:rho_0}
\end{equation}
respectively, where $v_{\rm F}$ is the Fermi velocity.

The other $g$ processes are written with the help of the boson 
representation of the right-going ($j=1$) and left-going ($j=2$) 
electron fields with spin $s$ ($s=\pm1$),
\begin{equation}
\psi_{js}(x)= 
\lim_{\alpha\rightarrow0}\frac1{\sqrt{2\pi\alpha}}
\exp\left[ \pm ik_{\rm F}x 
\mp i\frac1{\sqrt{2}}\left(   \phi_\rho(x)+s  \phi_\sigma(x) \right) 
+i\frac1{\sqrt{2}}\left( \theta_\rho(x)+s\theta_\sigma(x) \right) \right]
\;,
\ \ \ \ \ \ \mbox{for }j=\left\{\begin{array}{c} 1 \\ 2 \end{array}\right.
\;,
\end{equation}
where $\alpha$ is a cut-off parameter, $k_{\rm F}$ is the Fermi wave number, 
$\phi_\nu(x)$ and $\theta_\nu(x)$, 
$(\nu = \sigma, \rho)$, are the fields defined as
\begin{equation}
\left.\begin{array}{c} \phi_\nu(x) \\ \theta_\nu(x) \end{array}\right\}
= \mp \frac{i\pi}{L} \sum_{p(\neq0)} \frac1{p}
e^{-\alpha\mid p\mid/2 -ipx} \left[ \nu_1(p) \pm \nu_2(p) \right]
\;.
\end{equation}
The field $\Pi_\nu(x)$ defined as 
$\Pi_\nu(x)=\frac1\pi \partial_x \theta_\nu(x) $ is conjugate to 
the phase field $\phi_\nu(x)$,
\begin{equation}
\left[ \phi_\nu(x), \Pi_\mu(y) \right] = i \delta_{\nu,\mu} \delta(x-y)
\;.
\end{equation}
With no magnetic field and at half filling, 
a backward scattering between antiparallel spins ($g_{1\perp}$) and 
an Umklapp process ($g_3$) read
\begin{eqnarray}
H_{\sigma, {\rm int}} &=& 
g_{1\perp}\sum_s \int dx \ 
\psi_{1,s}^\dagger(x) \psi_{2,-s}^\dagger(x) 
\psi_{1,-s}(x) \psi_{2,s}(x) \nonumber \\
&=&
\frac{2g_{1\perp}}{(2\pi\alpha)^2} \int dx 
\cos(2\sqrt{2} \phi_\sigma(x))
\;,
\end{eqnarray}
\begin{eqnarray}
H_{\rho, {\rm int}} &=& 
g_3 \int dx \left(
\psi_{1,\uparrow}^\dagger(x) \psi_{1,\downarrow}^\dagger(x) 
\psi_{2,\downarrow}(x) \psi_{2,\uparrow}(x)
+ {\mbox h.c.} \right) \nonumber \\
&=&
\frac{2g_3}{(2\pi\alpha)^2} \int dx 
\cos(2\sqrt{2} \phi_\rho(x))
\;,
\end{eqnarray}
respectively. 
The bilinear terms are rewritten with the fields $\phi_\nu(x)$ and 
$\theta_\nu(x)$ as 
\begin{equation}
H_{\nu,0} = \int \frac{dx}{2\pi} 
\left[ u_\nu K_\nu \left( \partial_x \theta_\nu(x) \right)^2  +
\frac{u_\nu}{K_\nu} \left( \partial_x \phi_\nu(x) \right)^2 \right]
\;,
\end{equation}
with velocities, 
\begin{mathletters}
\begin{equation}
u_\sigma^2 = \left( v_{\rm F} - \frac{g_4}{2\pi} \right)^2 
- \left(\frac{g_{1\parallel}}{2\pi}\right)^2
\;,\label{eq:u_sigma}
\end{equation}
\begin{equation}
u_\rho^2 = \left( v_{\rm F} + \frac{g_4}{2\pi} \right)^2 
- \left(\frac{g_{1\parallel}-2g_2}{2\pi}\right)^2
\;,\label{eq:u_rho}
\end{equation}
\end{mathletters}\noindent
and correlation exponents,
\begin{mathletters}
\begin{equation}
K_\sigma^2 = \frac{ 2\pi v_{\rm F} - g_4 +g_{1\parallel} }
                  { 2\pi v_{\rm F} - g_4 -g_{1\parallel} }
\;,
\end{equation}
\begin{equation}
K_\rho^2 = \frac{ 2\pi v_{\rm F} + g_4 +g_{1\parallel}-2g_2 }
                { 2\pi v_{\rm F} + g_4 -g_{1\parallel}+2g_2 }
\;.
\end{equation}
\end{mathletters}\noindent

With a magnetic field $h$ or away from half filling, we add
\begin{equation}
  h \int dx \frac{\sqrt{2}}{\pi} \partial_x \phi_\sigma(x) 
+ \mu \int dx \frac{\sqrt{2}}{\pi} \partial_x \phi_\rho(x)
\;,
\end{equation}
where $\mu$ is the chemical potential set to be zero at half filling.
In order to work with fixed magnetization ($m$) and filling ($\delta n$), 
where $m=\pm1$ at saturated magnetization, $\delta n=0$ at half filling, 
and $\delta n=1$ at complete filling,
Legendre transforms are performed.\cite{Giamarchi}
Defining 
\begin{mathletters}
\begin{equation}
\psi_\sigma(x) = \phi_\sigma(x) + \frac{\pi}{\sqrt{2}} m x
\;,
\end{equation}
\begin{equation}
\psi_\rho(x) = \phi_\rho(x) + \frac{\pi}{\sqrt{2}} \delta n x
\;,
\end{equation}
\end{mathletters}\noindent
and rewriting $\psi_\sigma(x)$ and $\psi_\rho(x)$ as 
$\phi_\sigma(x)$ and $\phi_\rho(x)$, respectively, 
we get the electronic part of the model. 
The spin part, 
\begin{equation}
H_\sigma = H_{0,\sigma} + H_{1\perp}
\;,
\end{equation}
consists of the ``noninteracting'' part,
\begin{equation}
H_{0,\sigma} = \int \frac{dx}{2\pi} 
\left[ u_\sigma K_\sigma \left( \partial_x \theta_\sigma(x) \right)^2  +
\frac{u_\sigma}{K_\sigma} \left( \partial_x \phi_\sigma(x) \right)^2 +
\frac{\pi^2 u_\sigma m^2}{2K_\sigma} \right]
\;,
\end{equation}
and the perturbation due to the backward scattering,
\begin{equation}
H_{1\perp} = \frac{2g_{1\perp}}{(2\pi\alpha)^2} \int dx 
\cos(2\sqrt{2} \phi_\sigma(x)-2\pi mx)
\;,\label{eq:1perp}
\end{equation}
while the charge part,
\begin{equation}
H_\rho = H_{0,\rho} + H_{3}
\;,
\end{equation}
consists of the ``noninteracting'' part,
\begin{equation}
H_{0,\rho} = \int \frac{dx}{2\pi} 
\left[ u_\rho K_\rho \left( \partial_x \theta_\rho(x) \right)^2  +
\frac{u_\rho}{K_\rho} \left( \partial_x \phi_\rho(x) \right)^2 + 
\frac{\pi^2 u_\rho \delta n^2}{2K_\rho} \right]
\;,
\end{equation}
and the perturbation due to the Umklapp process,
\begin{equation}
H_{3} = \frac{2g_3}{(2\pi\alpha)^2} \int dx 
\cos(2\sqrt{2} \phi_\rho(x)-2\pi \delta nx)
\;.\label{eq:3}
\end{equation}
Here, we do not take the filling dependence of the Fermi velocity 
into account.  
In the lowest-order perturbation, the 
Hubbard model with the on-site interaction $U$ corresponds to 
$g_{1\parallel} = g_{1\perp} = g_2 = g_3 = g_4 = U$.\cite{Emery}

The phonon field is decomposed into a part with momenta near zero, 
$\phi_0(p)$, and the rest with momenta near $\pi$ (in units of 
the inverse lattice spacing), $\phi_\pi(x)$.
The former causes a forward scattering and is represented in the 
momentum space.  
The latter causes a backward scattering and is represented in the 
coordinate space.
This decomposition is therefore made possible by the decomposition 
of the electron field into the right- and left-going ones so that 
it is achieved unambiguously.
Their conjugate momenta are denoted by $\Pi_0(p)$ and $\Pi_\pi(x)$, 
respectively.
\begin{equation}
\left[ \phi_0(p), \Pi_0(q) \right] = i \delta_{p,q}
\;,
\end{equation}
\begin{equation}
\left[ \phi_\pi(x), \Pi_\pi(y) \right] = i \delta(x-y)
\;.
\end{equation}
Note that $\phi_\pi(x)$ and $\Pi_\pi(x)$ are real fields.

In the phonon part, we take dispersionless phonons for simplicity.
Later we will consider electron-phonon couplings in the Holstein 
model and in the SSH model when evaluating correlation exponents.
The neglect of the dispersion of bare phonons is not essential 
since it is absent in the former model and the acoustic phonons 
are decoupled from the low-energy electronic spectrum in the 
continuum limit of the latter model.\cite{Hirsch}
We divide the phonon part into a part with momenta near $\pi$ 
and the rest with momenta near zero,
\begin{equation}
H_{\rm ph} = \frac12 \int dx \left[ \Pi_\pi^2(x) 
+ \omega_\pi^2 \phi_\pi^2(x) \right]
+\frac1{2L} \sum_p \left[ \mid \Pi_0(p) \mid^2 
+ \omega_0^2 \mid \phi_0(p) \mid^2 \right]
\;,
\end{equation}
where $\omega_\pi$ and $\omega_0$ are their respective phonon frequencies.

The Holstein coupling to the dispersionless phonons is represented by 
\begin{equation}
\sum_i \left( \beta q_i n_i + \frac{K}2 q_i^2 + \frac1{2M} p_i^2 \right)
\;,\label{eq:Holstein}
\end{equation}
with the electron density at site $i$, $n_i$, the lattice displacement 
$q_i$, its conjugate momentum $p_i$, the coupling strength $\beta$, 
the spring constant $K$, and the ionic mass $M$.
In the continuum limit, it can be written as $H_1+H_2+H_{\rm ph}$, where 
$H_1$ denotes a phonon-assisted backward scattering,
\begin{equation}
H_1 = \gamma_1 \sum_s \int dx 
\left[ 
\psi_{2,s}^\dagger(x) \psi_{1,s}(x) e^{ i\pi \delta nx} +
\psi_{1,s}^\dagger(x) \psi_{2,s}(x) e^{-i\pi \delta nx}
\right] \phi_\pi(x)
\;,
\end{equation}
and $H_2$ stands for a phonon-assisted forward scattering,
\begin{equation}
H_2 = \gamma_2 \frac{\sqrt{2}}{L} \sum_p 
\left[ \rho_1(-p) + \rho_2(-p) \right] \phi_0(p)
\;,
\end{equation}
with $\gamma_1 = \gamma_2 = \beta / \sqrt{M}$ and 
$\omega_\pi = \omega_0 = \sqrt{K/M}$.
Note that the operators $\psi_{1,s}(x)$, $\psi_{2,s}(x)$, and $\phi_\pi(x)$
above concern with the slowly varying part, whereas the rapidly oscillating 
parts give the factors $e^{\pm i\pi \delta nx}=e^{\pm i(2k_{\rm F}-\pi)x}$.

The SSH coupling to the phonons is represented by
\begin{equation}
\sum_i \left[ \alpha_{\rm s} (q_{i+1}-q_i) \sum_s 
( c_{i,s}^\dagger c_{i+1,s} + c_{i+1,s}^\dagger c_{i,s} ) +
\frac{K}{2} (q_{i+1}-q_i)^2 + \frac1{2M} p_i^2
\right]
\;,\label{eq:SSH}
\end{equation}
with $c_{i,s}$ annihilating an electron with spin $s$ at site $i$, 
$\alpha_{\rm s}$ being the coupling strength, and $q_i$, $p_i$, $K$ and $M$ 
as above.
In the continuum limit, by neglecting the phonon dispersion,
it can be written as $H_{-1}+H_{\rm ph}$, where $H_{-1}$ denotes 
a phonon-assisted backward scattering, 
\begin{equation}
H_{-1} = \gamma_{-1} \sum_s \int dx 
\left[ 
 i \psi_{2,s}^\dagger(x) \psi_{1,s}(x) e^{ i\pi \delta nx} 
-i \psi_{1,s}^\dagger(x) \psi_{2,s}(x) e^{-i\pi \delta nx}
\right] \phi_\pi(x)
\;,
\end{equation}
with $\gamma_{-1} = 4\alpha_{\rm s}/\sqrt{M}$ and $\omega_\pi = 2\sqrt{K/M}$.
The scatterings $H_1$ and $H_{-1}$ differ in the form factor.

For later convenience, we define parameters $X_\nu$ depending only on 
$K_\nu$ and dimensionless coupling strengths $Y_\nu$ as\cite{Voit}
\begin{equation}
X_\nu = 2(1-K_\nu^{-1})
\:,
\end{equation}
\begin{mathletters}
\begin{equation}
Y_\sigma = g_{1\perp} /(\pi v_{\rm F})
\;,\label{eq:y_sigma}
\end{equation}
\begin{equation}
Y_\rho  = g_3 /(\pi v_{\rm F})
\;.\label{eq:y_rho}
\end{equation}
\end{mathletters}\noindent
In the lowest order, the parameters $X_\sigma$ and $X_\rho$ are 
given by
\begin{mathletters}
\begin{equation}
X_\sigma \simeq g_{1\parallel} /(\pi v_{\rm F})
\;,\label{eq:x_sigma}
\end{equation}
\begin{equation}
X_\rho  \simeq (g_{1\parallel} - 2 g_2)/(\pi v_{\rm F})
\;,\label{eq:x_rho}
\end{equation}
\end{mathletters}\noindent
respectively.
Hereafter, the Hubbard model is meant for 
$H_{\rm H} = H_\sigma + H_\rho$ with 
$X_\sigma = Y_\sigma = - X_\rho = Y_\rho \equiv Y_{\rm el} = 
U/(\pi v_{\rm F}) > 0$, 
the Holstein-Hubbard model for 
$H_{\rm H-H} = H_{\rm H} + H_{\rm ph} + H_1 + H_2$ with 
$\omega_\pi = \omega_0 \equiv \omega = \sqrt{K/M}$ and 
$\gamma_1 = \gamma_2 = \beta / \sqrt{M}$, and 
the SSH-Hubbard model for
$H_{\rm SSH-H} = H_{\rm H} + H_{\rm ph} + H_{-1}$ with 
$\omega_\pi \equiv \omega = 2\sqrt{K/M}$ and 
$\gamma_{-1} = 4\alpha_{\rm s}/\sqrt{M}$.
Numerical evaluations are performed also in the SSH-Hubbard model 
supplemented with a forward scattering,
$H_{\rm SSH-H-f.s.} = H_{\rm SSH-H} + H_2$, with 
$\omega_0 = \omega_\pi$ and $\gamma_2 = \gamma_{-1}$.

Effective interactions are obtained through the integration of phonon 
fields.
First we consider the antiadiabatic limit ($\omega\rightarrow\infty$), 
where the interactions are instantaneous.
In this limit, Eq.~(\ref{eq:Holstein}) becomes 
\begin{equation}
- \frac{\beta^2}{K} \sum_i 
c_{i\uparrow}^\dagger c_{i\uparrow} c_{i\downarrow}^\dagger c_{i\downarrow}
\;,
\end{equation}
which corresponds to 
$g_{1\parallel} = g_{1\perp} = g_2 = g_3 = g_4 = -\beta^2/K$, and 
Eq.~(\ref{eq:SSH}) becomes 
\begin{equation}
- \frac{\alpha_{\rm s}^2}{2K} \sum_{i,s,s'} 
(c_{i,s}^\dagger c_{i+1,s} + c_{i+1,s}^\dagger c_{i,s})
(c_{i,s'}^\dagger c_{i+1,s'} + c_{i+1,s'}^\dagger c_{i,s'})
\;,
\end{equation}
which corresponds to 
$g_{1\parallel} = g_{1\perp} = - g_3 = -4\alpha_{\rm s}^2/K$ 
and $g_2 = g_4 = 0$.
Later we will take either a phonon-assisted backward scattering originating 
from the on-site coupling ($\gamma_1$) or that from the intersite 
coupling ($\gamma_{-1}$), but not both.
We then define a dimensionless coupling strength $Y_1$ as 
\begin{equation}
Y_1 = \frac{\gamma_{\pm1}^2}{\pi v_{\rm F} \omega_\pi^2} 
\;.
\end{equation}
If necessary, the two cases above are distinguished by 
$f=1$ for $\gamma_1$ and $f=-1$ for $\gamma_{-1}$.
Another dimensionless coupling strength $Y_2$ is defined as 
\begin{equation}
Y_2 = \frac{\gamma_2^2}{\pi v_{\rm F} \omega_0^2}
\;.
\end{equation}
The subscripts of $Y_1$ and $Y_2$ stand for back and forward 
scatterings, respectively, as those in $g$ parameters.
In the Holstein-Hubbard model and the SSH-Hubbard model supplemented 
with a forward scattering, we define $Y_{\rm ph}$ as 
$Y_1 = Y_2 \equiv Y_{\rm ph}$.
Summarizing the antiadiabatic case, electron-phonon interaction shifts 
the electronic parameters by
$X_\sigma \rightarrow X_\sigma - Y_1$, $Y_\sigma \rightarrow Y_\sigma - Y_1$, 
$X_\rho \rightarrow X_\rho - (Y_1 - 2 Y_2)$, and $Y_\rho \rightarrow Y_\rho 
- f Y_1$.
From the definitions, (\ref{eq:y_sigma}), (\ref{eq:x_sigma}), and 
(\ref{eq:x_rho}), the first three are interpreted as due to effective 
attraction.  
The difference between the on-site coupling ($f=1$) and the intersite 
coupling ($f=-1$) appears in the effect on $Y_\rho$.
Its origin is easily understood in the continuum limit.
The effective $g_1$ processes come from the contraction of 
$\psi_{2,s}^\dagger(x)  \psi_{1,s}(x)$ and 
$\psi_{1,s'}^\dagger(x) \psi_{2,s'}(x)$ in $H_1$ or $H_{-1}$, 
whereas the effective $g_3$ processes come from the contraction of 
$\psi_{2,s}^\dagger(x)  \psi_{1,s}(x)$ and 
$\psi_{2,-s}^\dagger(x) \psi_{1,-s}(x)$ and that of 
$\psi_{1,s}^\dagger(x)  \psi_{2,s}(x)$ and 
$\psi_{1,-s}^\dagger(x) \psi_{2,-s}(x)$,
giving the sign factor $f$.
Hereafter, we will not consider a $g_4$ process, which only yields a 
renormalization of the Fermi velocity.

To treat finite phonon frequencies, we consider the partition function 
$Z = {\rm Tr} e^{-\beta H}$ at the inverse temperature $\beta$ and 
map the present system to a two-dimensional classical system 
in the Euclidean space ${\bf r}=(x,y=v_{\rm F}\tau)$ with an imaginary time 
$\tau$.\cite{Voit}
Because the phonon fields are bilinear in the Hamiltonian, 
they are analytically integrated out.
Effective interactions $H_{i,{\rm eff}}$ ($i=\pm1,2$) are defined as
\begin{equation}
{\rm T}_\tau \exp\left[ -\int_0^\beta d\tau H_{i,{\rm eff}}(\tau) \right]
= \left\langle {\rm T}_\tau \exp\left[ -\int_0^\beta d\tau H_i(\tau) \right] 
\right\rangle_{\rm ph}
\;,
\end{equation}
where T$_\tau$ is the time-ordering operator in the imaginary time, and 
$\langle \cdots \rangle_{\rm ph}$ denotes an expectation value 
averaged over phonon fields.
It is straightforward to derive, in the $\beta\rightarrow\infty$ limit, 
\begin{eqnarray}
H_{1,{\rm eff}}(\tau) &=& - \frac{\gamma_1^2}{\omega_\pi^2 (2\pi\alpha)^2}
\sum_{\epsilon_1=\pm1} \int dx \int d\Delta\tau 
\left[ \frac{\omega_\pi}{2} e^{-\omega_\pi \mid \Delta\tau \mid} \right]
\frac12 \sum_{\epsilon_2,\epsilon'_1,\epsilon'_2=\pm1}  
\exp\left\{
 i \epsilon_1 \left(\sqrt{2}\phi_\sigma(1)-\pi m x\right)
\right.\nonumber\\ & & \left.
-i \epsilon_2 \left(\sqrt{2}\phi_\sigma(2)-\pi m x\right)
+i \epsilon'_1\left(\sqrt{2}\phi_\rho(1)-\pi \delta nx\right)
-i \epsilon'_2\left(\sqrt{2}\phi_\rho(2)-\pi \delta nx\right)
\right\}
\;,\label{eq:1_eff}
\end{eqnarray}
\begin{equation}
H_{2,{\rm eff}}(\tau) = - \frac{2\gamma_2^2}{\omega_0^2} \frac1{L}
\sum_{p>0} \int d\Delta\tau 
\left[ \frac{\omega_0}{2} e^{-\omega_0 \mid \Delta\tau \mid} \right]
\left\{ \rho_1( p,\tau_1)+\rho_2( p,\tau_1) \right\}
\left\{ \rho_1(-p,\tau_2)+\rho_2(-p,\tau_2) \right\}
\;,\label{eq:2_eff}
\end{equation}
\begin{eqnarray}
H_{-1,{\rm eff}}(\tau) &=& - \frac{\gamma_{-1}^2}{\omega_\pi^2 (2\pi\alpha)^2}
\sum_{\epsilon_1=\pm1} \int dx \int d\Delta\tau 
\left[ \frac{\omega_\pi}{2} e^{-\omega_\pi \mid \Delta\tau \mid} \right]
\frac12 \sum_{\epsilon_2,\epsilon'_1,\epsilon'_2=\pm1}  
{\rm sgn}(\epsilon'_1 \epsilon'_2)
\exp\left\{
 i \epsilon_1 \left(\sqrt{2}\phi_\sigma(1)-\pi m x\right)
\right.\nonumber\\ & & \left.
-i \epsilon_2 \left(\sqrt{2}\phi_\sigma(2)-\pi m x\right)
+i \epsilon'_1\left(\sqrt{2}\phi_\rho(1)-\pi \delta nx\right)
-i \epsilon'_2\left(\sqrt{2}\phi_\rho(2)-\pi \delta nx\right)
\right\}
\;,\label{eq:-1_eff}
\end{eqnarray}
where arguments 1 and 2 denote $(x,v_{\rm F}\tau_1)$ and 
$(x,v_{\rm F}\tau_2)$, respectively, $\tau = (\tau_1 + \tau_2)/2$, 
and $\Delta\tau = \tau_1 - \tau_2$.
In the $\omega\rightarrow\infty$ limit, $[ \cdots ]$ above 
becomes $\delta(\tau_1 - \tau_2)$ so that the 
$  \epsilon_1 \epsilon_2 =   \epsilon'_1 \epsilon'_2 = 1$ terms 
in Eqs.~(\ref{eq:1_eff}) and (\ref{eq:-1_eff}) correspond to the 
$g_{1\parallel}$ process, 
$- \epsilon_1 \epsilon_2 =   \epsilon'_1 \epsilon'_2 = 1$ to the 
$g_{1\perp}$ process, 
$  \epsilon_1 \epsilon_2 = - \epsilon'_1 \epsilon'_2 = 1$ to the 
$g_3(=g_{3\perp})$ process, and 
$- \epsilon_1 \epsilon_2 = - \epsilon'_1 \epsilon'_2 = 1$ to the 
$g_{3\parallel}$ process not considered here, whereas the cross terms in 
Eq.~(\ref{eq:2_eff}) correspond to the $g_2(=g_{2\parallel}=g_{2\perp})$ 
processes, and the rest to the $g_{4\parallel}=g_{4\perp}$ processes.
Note that $\mp g_4 = \mp g_{4\perp}$ in Eqs.~(\ref{eq:sigma_0}) and 
(\ref{eq:rho_0}) are replaced by $g_{4\parallel} \mp g_{4\perp}$ if 
$g_{4\parallel} \neq 0$.
Again, we can derive 
$X_\sigma \rightarrow X_\sigma - Y_1$, $Y_\sigma \rightarrow Y_\sigma - Y_1$, 
$X_\rho \rightarrow X_\rho - (Y_1 - 2 Y_2)$, and $Y_\rho \rightarrow Y_\rho 
- f Y_1$ in this limit.

Finally, we have the ``noninteracting'' part, $H_0 = H_{0,\sigma} 
+ H_{0,\rho}$, and perturbations, $H_{1\perp}$, $H_3$, 
$H_{\pm1,{\rm eff}}(\tau)$, and $H_{2,{\rm eff}}(\tau)$.

\section{RENORMALIZATION EQUATIONS}\label{sec:renormalization}

Renormalization group equations are derived in a way similar to 
Voit-Schulz\cite{Voit} and Giamarchi-Schulz\cite{Giamarchi}.
Because of the similarity, we will not go into details but will 
give an outline detailed enough to explain a few modifications.
We consider the correlation function,
\begin{equation}
R_\nu({\bf r_{12}}) = \left\langle
{\rm T}_\tau 
e^{i\sqrt{2}\phi_\nu({\bf r_1})} e^{-i\sqrt{2}\phi_\nu({\bf r_2})}
\right\rangle
\;,
\end{equation}
with ${\bf r_{12}} = {\bf r_1} - {\bf r_2}$ and develop in powers of 
$Y_\sigma$, $Y_\rho$, $Y_1$, and $Y_2$.
In the zeroth order, it is given by 
\begin{equation}
R_\nu^{(0)}({\bf r_{12}}) = \exp\left(
-\frac{K_\nu}{2} \log \frac{r_{12}^2+\alpha^2}{\alpha^2}
\right)
\;.
\end{equation}
In the expansion, we sum only over those configurations which 
satisfy the spin- and charge-conservation conditions: 
the sum of the coefficients of the $\phi_\sigma$ terms in the exponents 
and that of the $\phi_\rho$ terms are zero.
Each term is written as a product of a spin part and a charge part 
the exponents of which are linear in $\phi_\sigma$ and $\phi_\rho$, 
respectively, so that the expectation value is easily obtained.
As a result, each term is written as an integral the variables of which 
can be regarded as positions 
of Burgers vectors in the Euclidean space\cite{Voit} 
and denoted by ${\bf r_i}$, $(i=a$, $b$, $\cdots)$.
The perturbation $H_{\pm1,{\rm eff}}(\tau)$ 
produces two vectors each of which is allowed to take one of four possible 
orientations $\pm {\bf b_1}$ and $\pm {\bf b_2}$ with 
${\bf b_{1,2}} = (K_\sigma^{1/2}, \pm K_\rho^{1/2})$.
The perturbation $H_{1\perp}$ produces a vector ${\bf b_\sigma}$ or  
$-{\bf b_\sigma}$ with ${\bf b_\sigma} = (2 K_\sigma^{1/2}, 0)$, while 
$H_3$ produces a vector ${\bf b_\rho}$ or $-{\bf b_\rho}$ with 
${\bf b_\rho} = (0, 2K_\rho^{1/2})$.
The perturbation $H_{2,{\rm eff}}(\tau)$ is represented with 
different operators and not interpreted as producing Burgers vectors: 
it will be treated later.
For example, the $O(Y_\nu^2)$ term, ($\nu = \sigma, \rho$), is 
regarded as arising from a logarithmic interaction, 
\begin{equation}
- {\bf b} ({\bf r_a}) \cdot {\bf b} ({\bf r_b}) \frac12 \log
\frac{\mid {\bf r_a} - {\bf r_b} \mid^2 + \alpha^2}{\alpha^2}
\;,
\end{equation}
with ${\bf b} ({\bf r_a}) = - {\bf b} ({\bf r_b}) = \pm {\bf b_\nu}$ 
in a classical system of Burgers vectors if $m=0$ and $\delta n=0$.
With $m\neq0$ and $\delta n\neq0$, the corresponding interaction 
becomes anisotropic in the Euclidean space.\cite{Giamarchi}
The $O(Y_1)$ term corresponds to a contribution from a 
logarithmic plus linear potential,
\begin{equation}
- {\bf b} ({\bf r_a}) \cdot {\bf b} ({\bf r_b}) \frac12 \log
\frac{\mid {\bf r_a} - {\bf r_b} \mid^2 + \alpha^2}{\alpha^2}
- \frac{\mid y_a - y_b \mid}{\xi_\pi}
\;,
\end{equation}
with a constraint $x_a = x_b$, 
${\bf b} ({\bf r_a}) = - {\bf b} ({\bf r_b}) = \pm {\bf b_i}$, ($i=1,2$), 
and $\xi_\pi = v_{\rm F}/\omega_\pi$.
The second term and the constraint come from the phonon propagator 
in Eqs.~(\ref{eq:1_eff}) and (\ref{eq:-1_eff}),
\begin{equation}
D_\pi({\bf r_a} - {\bf r_b}) = \frac1{2 \omega_\pi} 
\delta(x_a - x_b) \exp[-\frac{\mid y_a - y_b \mid}{\xi_\pi}]
\;,
\end{equation}
which becomes isotropic in the antiadiabatic limit.
The integral in each term is dominated by configurations with 
${\bf r_a}$ very close to ${\bf r_b}$.
Then it is convenient to change the integration variables to 
${\bf R} = ({\bf r_a} + {\bf r_b})/2$ and ${\bf r} = {\bf r_a} - {\bf r_b}$
and expand the integrand to the second order in ${\bf r}$.
Integration can be performed over ${\bf R}$.\cite{Giamarchi}
The remaining integration over ${\bf r}$ is cut off at $\alpha$ 
($r > \alpha$) and the function $\log[(r^2 + \alpha^2)/\alpha^2]/2$ is 
replaced by $\log(r/\alpha)$.

We will search a renormalized correlation function of the form,
\begin{equation}
\exp\left(
-\frac{K_\nu^{\rm eff}}{2} \log \frac{r_{12}^2+\alpha^2}{\alpha^2}
-d_\nu^{\rm eff} \frac{x_{12}^2-y_{12}^2}{r_{12}^2}
\right)
\;,
\end{equation}
regarding the corrections 
as modifying the exponent $K_\nu$ to an effective one $K_\nu^{\rm eff}$ 
and producing the anisotropy parameter $d_\nu^{\rm eff}$.
Changing the cut off $\alpha$ from $\alpha(l) \equiv \alpha e^l$ to 
$\alpha(l+dl)$, a pair of Burgers vectors with 
$\alpha(l) < r < \alpha(l+dl)$ are no more visible.
This portion 
of the integrals can be absorbed into changing $K_\nu(l) 
[=1/(1-X_\nu(l)/2)]$ and $d_\nu(l)$ into $K_\nu(l+dl)$ and $d_\nu(l+dl)$.
Upon rescaling the remaining integrals so that they again range from 
$\alpha$ to infinity, the effective parameters $K_\nu^{\rm eff}$ and 
$d_\nu^{\rm eff}$ are written with modified couplings $Y_\nu(l+dl)$ and 
$Y_1(l+dl)$.
In addition, nonneutral pairs of Burgers vectors with 
$\alpha(l) < r < \alpha(l+dl)$ 
appearing in the $O(Y_\nu Y_1)$, $O(Y_1^2)$, $O(Y_\nu^2 Y_1)$ terms, etc., 
are combined into new vectors:
${\bf b_1}$ and ${\bf b_2}$ into ${\bf b_\sigma}$, 
${\bf b_1}$ and $-{\bf b_2}$ into ${\bf b_\rho}$, 
${\bf b_\sigma}$ and $-{\bf b_2} (-{\bf b_1})$ into ${\bf b_1} ({\bf b_2})$, 
${\bf b_\rho}$ and ${\bf b_2} (-{\bf b_1})$ into ${\bf b_1} (-{\bf b_2})$, etc.
They also contribute to modifying $Y_\nu(l)$ and $Y_1(l)$ into 
$Y_\nu(l+dl)$ and $Y_1(l+dl)$.
The magnetic field $h$ and the chemical potential $\mu$ are evaluated 
perturbationally.
From the scaling $h(l) = e^l h$ and $\mu(l) = e^l \mu$,\cite{Giamarchi}
we derive the relation between $m(l)$ and $m(l+dl)$ and that between
$\delta n(l)$ and $\delta n(l+dl)$.
The factors $e^{\pm i2\pi mx}$ in Eq.~(\ref{eq:1perp}), 
$e^{\pm i2\pi \delta nx}$ in Eq.~(\ref{eq:3}), and 
$(\omega_\pi/2)e^{-\omega_\pi \mid \Delta\tau \mid}$ 
in Eqs.~(\ref{eq:1_eff}) and (\ref{eq:-1_eff}) make the correlation 
functions anisotropic in the Euclidean space.
A resultant change of the $y$ scale relative to the $x$ scale in 
$R_\nu({\bf r_{12}})$ can be interpreted as a change of the velocity in the 
$\nu$ channel:
$d[d_\nu(l)]/dl =  -[K_\nu(l)/2] d \log[u_\nu(l)]/dl$ in the lowest order.

Finally, we obtain the renormalization group equations,
\begin{equation}
d X_\sigma(l)/d l = - Y_\sigma^2(l) J_0(2\pi m(l) \alpha) - Y_1(l) D_\pi(l)
\;,\label{eq:x_sigma_l} 
\end{equation}
\begin{equation}
d Y_\sigma(l)/d l = - X_\sigma(l) Y_\sigma(l) - Y_1(l) D_\pi(l)
\;,\label{eq:y_sigma_l} 
\end{equation}
\begin{equation}
d (2\pi m(l) \alpha)/d l = 2\pi m(l) \alpha + Y_\sigma^2(l) 
J_1(2\pi m(l) \alpha) 
\;,\label{eq:m_l}
\end{equation}
\begin{equation}
d \log[u_\sigma(l)]/d l = -\frac12 K_\sigma(l) \left[
Y_\sigma^2(l) J_2(2\pi m(l) \alpha) + Y_1(l) D_\pi(l) \right]
\;,\label{eq:u_sigma_l}
\end{equation}
\begin{equation}
d X_\rho(l)/d l = - Y_\rho^2(l) J_0(2\pi \delta n(l) \alpha) - Y_1(l) D_\pi(l)
+2 Y_2(l) D_0(l)
\;,\label{eq:x_rho_l}
\end{equation}
\begin{equation}
d Y_\rho(l)/d l = - X_\rho(l) Y_\rho(l) - f Y_1(l) D_\pi(l)
\;,\label{eq:y_rho_l}
\end{equation}
\begin{equation}
d (2\pi \delta n(l) \alpha)/d l = 2\pi \delta n(l) \alpha + 
Y_\rho^2(l) J_1(2\pi \delta n(l) \alpha)
\;,\label{eq:n_l}
\end{equation}
\begin{equation}
d \log[u_\rho(l)]/d l = -\frac12 K_\rho(l) \left[
Y_\rho^2(l) J_2(2\pi \delta n(l) \alpha) + Y_1(l) D_\pi(l) \right]
\;,\label{eq:u_rho_l}
\end{equation}
\begin{equation}
d Y_1(l)/d l = \left[ 2 - K_\sigma(l) - K_\rho(l)
- Y_\sigma(l) J_0(2\pi m(l) \alpha) 
- f Y_\rho(l) J_0(2\pi \delta n(l) \alpha) \right] Y_1(l) 
\;,\label{eq:y_1_l}
\end{equation}
\begin{equation}
d Y_2(l)/d l = 0
\;,\label{eq:y_2_l}
\end{equation}
where $D_\mu(l) = [\alpha(l)/\xi_\mu]\exp[-\alpha(l)/\xi_\mu]$, 
($\mu = 0, \pi$).
Since the cut-off parameter $\alpha$ should be physically 
of the order of $k_{\rm F}^{-1}$,
$D_\mu(l)$ can be regarded as 
$D_\mu(l) = [\omega_\mu/E(l)]\exp[-\omega_\mu/E(l)]$, 
with $E(l) = E_{\rm F} e^{-l}$ and $E_{\rm F}$ being the Fermi energy.
The $J_n(x)$ are the Bessel functions, whose oscillations spuriously 
come from the use of a sharp cut off in real space.
In numerical integrations, we replace them by $J_n(x) 
\theta(j_{0,1}-\mid x \mid)$ with $j_{0,1}$ being the first zero of 
$J_0(x)$ [$J_0(j_{0,1}) = 0$], $\theta(x) = 1$ for $x>0$ and 
$\theta(x) = 0$ for $x<0$.
The first terms in Eqs.~(\ref{eq:y_sigma_l}) and (\ref{eq:y_rho_l}) 
are linearized in $X_\nu$ since we deal with the lowest-order perturbations.
In the absence of $Y_1$ and $Y_2$, these equations are reduced to 
known ones.\cite{Solyom}
Furthermore, they are consistent with the isotropy in spin space 
[$X_\sigma(l) = Y_\sigma(l)$ when $m=0$].
The velocities $u_\nu$ would be modified also through 
Eqs.~(\ref{eq:u_sigma}) and (\ref{eq:u_rho}), but it is of higher order 
and neglected here.
In the lowest order considered here, they are decoupled from the rest.
Compared with the previous works\cite{Voit,Voit_prl}, the factor 
$e^l$ is absorbed into the prefactor of $D_\mu(l)$ so that the right-hand 
sides of Eqs.~(\ref{eq:y_1_l}) and (\ref{eq:y_2_l}) are not 
$(3 - \cdots) Y_1(l)$ and $Y_2(l)$ any more, respectively.
If the function 
$\langle {\rm T}_\tau [\phi_\nu({\bf r_1}) - \phi_\nu({\bf r_2})]^2 \rangle$ 
were renormalized as in Ref.~\cite{Voit}, 
a factor of 2 would appear in the above equations.
To keep the equations similar to those in the references, we doubled the 
definition of $Y_1$.
In Eq.~(\ref{eq:x_rho_l}), the effect of a phonon-assisted forward scattering 
is twice as large as that of a phonon-assisted backward scattering, as 
is expected from Eq.~(\ref{eq:x_rho}), but it is not in Ref.~\cite{Voit_prl}.
As we show below, 
the present equations correctly reproduce the antiadiabatic limit discussed
before.

To reproduce the antiadiabatic limit, we should recall that 
$\log[(r^2 + \alpha^2)/\alpha^2]/2$ was replaced above by $\log(r/\alpha)$ 
before the renormalization procedure.
Because the largest electron-phonon contribution comes from $r \sim \xi_\mu$, 
($\mu = 0, \pi$), it would be valid if $\alpha$ were much smaller 
than $\xi_\mu$.
However, a large value of $\omega_\mu / E_{\rm F}$ corresponds to 
a large value of $\alpha / \xi_\mu$, so that this replacement is 
no longer valid.
Instead, we will derive the $Y_1$ and $Y_2$ terms in the equations for 
$X_\nu$ and $Y_\nu$, ($\nu = \sigma, \rho$), in a more straightforward 
manner.
Equations~(\ref{eq:1_eff}), (\ref{eq:2_eff}), and (\ref{eq:-1_eff}) 
for the effective couplings $H_{i,{\rm eff}}$, ($i = \pm1, 2$), 
represent the electron-phonon contribution to the effective $g$ parameters 
with a retardation factor 
$(\omega_\mu/2) \exp(-\omega_\mu \mid \Delta\tau \mid)$.
As the time scale changes from $E(l)^{-1}$ to $E(l+dl)^{-1}$, 
the integration of the above factor over the time slice with width 
$E(l)^{-1} dl$ can be absorbed into definitions of the $g$ parameters,
for example, as
\begin{mathletters}
\begin{eqnarray}
g_{1\parallel}(l+dl) &=& g_{1\parallel}(l) - 
\frac{\gamma_{\pm1}^2}{\omega_\pi^2}
\left( \int_{-E(l+dl)^{-1}}^{-E(l)^{-1}} d\Delta\tau 
 + \int_{E(l)^{-1}}^{E(l+dl)^{-1}} d\Delta\tau  \right)
 \frac{\omega_\pi}{2} e^{-\omega_\pi \mid \Delta\tau \mid} \nonumber \\
 &=& g_{1\parallel}(l) - \frac{\gamma_{\pm1}^2}{\omega_\pi^2}
 D_\pi(l) dl
\;,
\end{eqnarray}
and similarly as
\begin{equation}
g_{1\perp}(l+dl) = g_{1\perp}(l) - \frac{\gamma_{\pm1}^2}{\omega_\pi^2}
 D_\pi(l) dl
\;,
\end{equation}
\begin{equation}
g_{3}(l+dl) = g_{3}(l) - f \frac{\gamma_{\pm1}^2}{\omega_\pi^2}
 D_\pi(l) dl
\;,
\end{equation}
\begin{equation}
g_{2}(l+dl) = g_{2}(l) - \frac{\gamma_{2}^2}{\omega_0^2}
 D_0(l) dl
\;,
\end{equation}\label{eq:retardation}
\end{mathletters}

\noindent
and an equation for $g_4(l+dl)$ similar to that for $g_{2}(l+dl)$. 
Note that here we do not take account of the effective $g$ processes 
of the purely electronic origin.
From these equations, the $Y_1$ terms are derived again and 
the $Y_2$ term is derived here.
Since $H_{2,{\rm eff}}(\tau)$ does not produce Burgers vectors,
$Y_2$ is independent of the scale.
It simply modifies $X_\rho(l)$ without interference with other 
quantities.
Note that similar renormalization group equations can be 
derived with Feynman diagrams.
Equations~(\ref{eq:y_1_l}) and (\ref{eq:y_2_l}) are interpreted as 
originating from corrections to the electron-phonon vertices.
The absence of a correction to the $Y_2(l)$ term 
is due to the fact that the bubble diagram in the zero-sound channel 
appearing in corrections to the phonon-assisted backward scattering 
gives the logarithmic correction [Eq.~(\ref{eq:y_1_l})]
while that in the third channel appearing in corrections to the 
phonon-assisted forward scattering does not give a logarithmic 
correction [Eq.~(\ref{eq:y_2_l})].\cite{Solyom}

Returning to the discussion of the antiadiabatic case, we should note 
that the variable $l$ ranges from 0 to $\infty$ so that $E(l)$ from 
$E_{\rm F}$ to 0 in Eqs.~(\ref{eq:x_sigma_l}), 
(\ref{eq:y_sigma_l}), (\ref{eq:x_rho_l}), and (\ref{eq:y_rho_l}).
It corresponds to $ E_{\rm F}^{-1} < \mid \Delta\tau \mid < \infty $ 
in spite of unconstrained lattice dynamics at $0 < \mid \Delta\tau 
\mid < \infty$.
This is in contrast to the purely electronic processes contributing 
to the effective $g$ processes [the first terms in 
Eqs.~(\ref{eq:x_sigma_l}), (\ref{eq:y_sigma_l}), (\ref{eq:x_rho_l}), 
and (\ref{eq:y_rho_l})] and the corrections to the electron-phonon 
vertex [Eq.~(\ref{eq:y_1_l})], 
which are limited to $E_{\rm F} > E(l) > 0$
[not included in Eqs.~(\ref{eq:retardation})].
Therefore, we integrate the second and third terms in 
Eqs.~(\ref{eq:x_sigma_l}), (\ref{eq:y_sigma_l}), (\ref{eq:x_rho_l}), 
and (\ref{eq:y_rho_l}) over $-\infty < l < 0$ 
corresponding to $ 0 < \mid \Delta\tau \mid < E_{\rm F}^{-1} $ 
with fixed values of $Y_1$ and $Y_2$ 
before integrating the renormalization group equations over 
$0 < l < \infty$.
It leads us to choose the initial conditions as\cite{Voit_prl}
\begin{mathletters}
\begin{equation}
X_\sigma(0) = X_\sigma - Y_1 \left( 1 - e^{-\omega_\pi/E_{\rm F}} \right)
\;,
\end{equation}
\begin{equation}
Y_\sigma(0) = Y_\sigma - Y_1 \left( 1 - e^{-\omega_\pi/E_{\rm F}} \right)
\;,
\end{equation}
\begin{equation}
X_\rho(0) = X_\rho - Y_1 \left( 1 - e^{-\omega_\pi/E_{\rm F}} \right)
+ 2 Y_2 \left( 1 - e^{-\omega_0/E_{\rm F}} \right)
\;,
\end{equation}
\begin{equation}
Y_\rho(0) = Y_\rho - f Y_1 \left( 1 - e^{-\omega_\pi/E_{\rm F}} \right)
\;.
\end{equation}
\end{mathletters}\noindent
In the antiadiabatic limit, the $Y_1$ and $Y_2$ terms are dropped from 
the renormalization group equations so that electron-phonon interaction 
simply shifts the initial conditions as discussed in Sec.~\ref{sec:model}.

\section{RESULTS}\label{sec:results}

First we consider analytic properties of the equations.
The phonon propagator $D_\mu(l)$ has a maximum at $E(l) = \omega_\mu$ 
and decreases exponentially for $E(l) < \omega_\mu$.
This smoothly cuts off the contribution from electron-phonon 
interaction at energy of the order of phonon frequency.
Since $Y_1$ also changes exponentially, 
the actual cut off is different from $\omega_\pi$.
Except for unphysically strong, intersite electron-phonon coupling 
(in view of the fact that 
$\alpha_{\rm s} (q_{i+1}-q_i)$ must be smaller than the bare 
transfer integral in the SSH model), $Y_1(l) D_\pi(l)$ finally 
decreases exponentially at large $l$ as well as $Y_2(l) D_0(l)$.
Then the $Y_1$ and $Y_2$ terms become negligible so that the equations 
for the spin part and for the charge part are decoupled to be  
in the familiar forms: effectively electron-phonon coupling is no more 
retarded and the system becomes equivalent to a purely 
electronic system with modified coupling strengths.

It should be noted that a spin-charge coupling at finite energy 
is a general phenomenon regardless of the presence of 
electron-phonon interaction: 
spin and charge are separated (in general only) asymptotically, 
i.e., in the limit of low energy and long wave length.
Some purely electronic interaction may induce a filling dependence 
of the asymptotic spin correlation through a spin-charge coupling and 
an Umklapp process at finite energy.
However, the Hubbard model is not the case since the fixed point of 
$K_\sigma$, $K_\sigma^\ast$, is known to be unity for $U>0$ 
regardless of filling.
Recall that, if both of the spin part and the charge part belong to the
Tomonaga-Luttinger class, 
the charge-density-wave correlation is proportional 
to $r^{-K_\sigma^\ast -K_\rho^\ast}$, 
the spin-density-wave correlation 
to $r^{-K_\sigma^{\ast-1} -K_\rho^\ast}$, 
the singlet-superconductor correlation 
to $r^{-K_\sigma^\ast -K_\rho^{\ast-1}}$, and 
the triplet-superconductor correlation 
to $r^{-K_\sigma^{\ast-1} -K_\rho^{\ast-1}}$,
in the asymptotic limit. 

Deviation from half filling and nonzero magnetization, if any, grow 
exponentially with $l$.
If a charge density wave is formed at half filling, 
charged solitons would be formed upon doping to connect two 
degenerate phases.
The quantity $\delta n \alpha$ can be regarded as the expected 
number of charged solitons inside the length $\alpha$.
As the renormalization proceeds with increasing length scale 
and decreasing energy scale, the Umklapp process $Y_\rho$ becomes 
ineffective at finite energy which increases with doping.
If magnetization exists, a similar argument holds: 
the backward scattering $Y_\sigma$ becomes 
ineffective at finite energy which increases with magnetization.
Then both terms on the right-hand side of Eq.~(\ref{eq:x_sigma_l}) are 
finally integrated out to make $X_\sigma(l)$ converge to some value:  
$K_\sigma(l)$ does not vanish so that the spin excitation spectrum is 
gapless.

Hereafter we always assume that magnetization is zero.
Then, for spin-independent coupling, $X_\sigma = Y_\sigma$, 
as in the Hubbard model, 
the two quantities are equal at any energy scale, 
$X_\sigma(l) = Y_\sigma(l)$.
After electron-phonon interaction is integrated out ($l>l_0$), 
it becomes of the form, $X_\sigma(l) = [l-l_0+X_\sigma^{-1}(l_0)]^{-1}$.
There are two possible fixed points: $X_\sigma^\ast = 0$ and 
$K_\sigma^\ast = 1$ if $X_\sigma(l_0)$ is positive; 
$X_\sigma^\ast = -\infty$ and $K_\sigma^\ast = 0$ 
if $X_\sigma(l_0)$ is negative.
In the former, the asymptotic spin correlation is the same as 
in the noninteracting case ($g_{1\parallel}=0$).
In the latter, the spin part corresponds to the Luther-Emery 
state\cite{Luther} so that the spin excitation spectrum has a gap.
For $\omega = \infty$, the spin gap opens if 
$Y_{\rm ph} > Y_{\rm el}$.
For $Y_{\rm el} = 0$, the spin gap opens whenever 
$Y_{\rm ph} > 0$ regardless of the phonon frequency 
since $X_\sigma(0)$ is negative and 
$X_\sigma(l)$ decreases at an increasing rate.

At half filling, the Umklapp process remains effective unless 
it is renormalized zero.
After electron-phonon interaction is integrated out ($l>l_0$), 
$X_\rho(l)$ and $Y_\rho(l)$ flow on a line 
$X_\rho^2(l) - Y_\rho^2(l) = X_\rho^2(l_0) - Y_\rho^2(l_0)$ 
in the direction of decreasing $X_\rho(l)$.
If $X_\rho(l_0) \geq \mid Y_\rho(l_0) \mid$, the fixed point has 
$X_\rho^\ast = \sqrt{X_\rho^2(l_0) - Y_\rho^2(l_0)}$ and $Y_\rho^\ast = 0$ 
so that $K_\rho^\ast \geq 1$.
The superconductor correlation is dominant if $K_\rho^\ast > 1$.
In the Holstein-Hubbard model with 
$Y_{\rm ph} > Y_{\rm el}$ and $\omega = \infty$, 
the initial conditions 
$X_\sigma(0) = Y_\sigma(0) = -X_\rho(0) = Y_\rho(0) = 
Y_{\rm el} - Y_{\rm ph}$ lead to 
$K_\sigma^\ast = 0$ and $K_\rho^\ast = 1$ so that it has a spin gap 
and degenerate singlet-superconductor and charge-density-wave 
correlations.
It is consistent with known results for the equivalent system, namely, 
the attractive Hubbard model.
If $X_\rho(l_0) < \mid Y_\rho(l_0) \mid$ on the other hand, 
$X_\rho(l)$ flows to $-\infty$ so that $K_\rho^\ast =0$ and the 
charge excitation spectrum has a gap.
This happens quite generally.
For $\omega = \infty$, it applies to the SSH-Hubbard model regardless of 
supplementation with a forward scattering and to the Holstein-Hubbard model
if  $Y_{\rm el} > Y_{\rm ph}$.
For the SSH-Hubbard model ($f=-1$ and $Y_2=0$), the relation 
$X_\rho(l) = - Y_\rho(l)$ holds at any energy scale: the charge gap 
always opens regardless of the phonon frequency 
since $X_\rho(0)$ is negative and 
$X_\rho(l)$ decreases at an increasing rate.
For the Holstein-Hubbard model ($f=1$ and $Y_2=Y_1$), the charge gap 
appears to open as is indeed numerically supported 
whenever the phonon frequency is finite: 
the initial condition is on the line $X_\rho(0) = - Y_\rho(0)$ and 
the electron-phonon interaction deviates the flow to the 
$X_\rho(0) + Y_\rho(0) < 0$ region even if $X_\rho(0) > 0 > Y_\rho(0)$
because $Y_1(l)$ initially increases and overcomes the effect of $Y_2(l)$ then.

Consequently, three phases are possible at half filing:
a charge-density-wave phase with both spin and charge gaps (SCG-CDW)
with $K_\sigma^\ast = 0$ and $K_\rho^\ast = 0$; 
a charge-gap phase with a dominant density-wave correlation (CG-DW) 
with $K_\sigma^\ast = 1$ and $K_\rho^\ast = 0$; and a spin-gap phase with 
degenerate singlet-superconductor and charge-density-wave correlations (SG)
with $K_\sigma^\ast = 0$ and $K_\rho^\ast = 1$.
The above facts of the spin and charge correlations at half filling 
are consistent with analytical and quantum Monte Carlo results 
for $Y_{\rm el} = 0$:\cite{Hirsch}  
the SCG-CDW phase always appears except for the Holstein model 
in the antiadiabatic limit where the SG phase appears instead.
The SCG-CDW phase in the Holstein model has an ordinary charge density wave, 
while that in the SSH model has a bond order wave, where bond lengths and 
bond charge density are alternated but the charge density is uniform.
Both cases have a similarity to a band insulator in the sense that 
they have both the spin and charge gaps.
On the other hand, in the CG-DW phase, only the charge gap opens.
It is similar to the repulsive Hubbard model.
Without a logarithmic correction to the correlation functions, 
it is not clear whether 
the charge-density-wave order is dominant or 
the spin-density-wave order is dominant.

Phase diagrams at half filling are shown in Fig.~\ref{fig:half_filling}.
With $Y_{\rm el} > 0$, $Y_{\rm ph}$ must be larger than a critical 
coupling strength to have the SCG-CDW phase.
The effect of electron-phonon interaction on a spin gap 
depends sensitively on whether the coupling is on-site or intersite, 
especially when the phonon frequency is low: 
the critical coupling strength is much lower for the intersite coupling 
than for the on-site coupling.
This is due to the sign factor $f$ in the correction to the electron-phonon 
vertex, Eq.~(\ref{eq:y_1_l}), and in the correction to the effective 
Umklapp process, Eq.~(\ref{eq:y_rho_l}).
For the intersite coupling ($f=-1$), the increasing Umklapp process 
[$Y_\rho(l)$] strengthens the phonon-assisted backward scattering [$Y_1(l)$], 
contrary to the on-site coupling ($f=1$).
The Umklapp process is further strengthened 
by the phonon-assisted backward scattering: these two processes are 
constructively interfered with each other.
The growing $Y_1(l)$ helps the opening of a spin gap [Eq.~(\ref{eq:x_sigma_l})] 
and the growth of the density-wave correlation [Eq.~(\ref{eq:x_rho_l})].
As shown later, this difference is evident near half filling also: 
the intersite coupling favors the spin gap and the density-wave correlation.
It is reasonable in view of the fact that, 
in the limit of strong on-site repulsion 
and small phonon frequency, 
the intersite coupling enables alternation of the superexchange 
and produces a spin-Peierls state no matter how the coupling is small.
Meanwhile, the on-site coupling must be large enough to compete 
with the on-site repulsion to affect the system.
As the phonon frequency increases, however, the difference between these two 
couplings becomes small. 
In the antiadiabatic limit, the spin gap opens for  
$Y_{\rm ph} > Y_{\rm el}$ in both cases.
As a function of the phonon frequency $\omega$, the critical coupling strength 
increases with $\omega$ for the intersite coupling, while it decreases for 
the one-site coupling.

Away from half filling, the fixed point $K_\rho^\ast$ is always positive 
and the charge excitation spectrum is gapless since the Umklapp process 
is integrated out at finite energy.
Therefore, four phases are possible: 
a gapless phase with a dominant superconductor correlation (SC) 
with $K_\sigma^\ast = 1$ and $K_\rho^\ast > 1$;
a gapless phase with a dominant density-wave correlation (DW) 
with $K_\sigma^\ast = 1$ and $K_\rho^\ast < 1$;
a singlet superconductor phase with a spin gap (SG-SC) 
with $K_\sigma^\ast = 0$ and $K_\rho^\ast > 1$; and 
a phase with a spin gap and a dominant 
charge-density-wave correlation (SG-CDW) 
with $K_\sigma^\ast = 0$ and $K_\rho^\ast < 1$.
The DW and SG-CDW phases have a finite Drude weight, 
$2 K_\rho^\ast u_\rho^\ast$, so that they are metallic.
The DW phase appears in the doped repulsive Hubbard model.
The SG-SC phase is regarded as a conventional superconductor phase since it is 
accompanied by a spin gap with the help of electron-phonon interaction, 
although a true long-range order is absent.

Far away from half filling where the Umklapp process is ineffective, 
a phase diagram is divided basically into 
the DW phase for $Y_{\rm el} \gsim Y_{\rm ph}$ and 
the SG-SC phase for $Y_{\rm ph} \gsim Y_{\rm el}$ 
(Figs.~\ref{fig:HH_phase_diagram} and 
\ref{fig:SHfs_phase_diagram}).
For finite phonon frequency, the SC phase occupies 
a narrow region between the two [not shown in 
Figs.~\ref{fig:HH_phase_diagram} (a) and 
\ref{fig:SHfs_phase_diagram} (c)] and may be 
an artifact of the lowest-order renormalization group approach.
However, the spin correlation and the charge correlation 
are separated in the asymptotic limit so that there is no reason to 
exclude the SC phase.
A phonon-assisted backward scattering tends to open a spin gap, 
while it enhances the density-wave correlation. 
To make the superconductor correlation dominant, a phonon-assisted 
forward scattering must overcome the backward scattering.
In the continuum limit of the SSH-Hubbard model where a coupling with 
acoustic phonons is absent, the SG-SC phase is 
made possible by the supplementation with a forward scattering.\cite{Voit_prl}

Phase diagrams for the Holstein-Hubbard model are shown in 
Fig.~\ref{fig:HH_phase_diagram}.
For very small phonon frequency $\omega$ and near half filling, 
the SG-CDW phase 
appears between the DW and SG-SC phases.
In Fig.~\ref{fig:HH_phase_diagram} (a), as the filling approaches 
the half ($\delta n=0$), the boundary between the SG-CDW and DW phases 
goes upwards abruptly in a very close vicinity of the half filling, although 
it occurs too close to distinguish the boundary from the ordinate. 
With $Y_{\rm ph}$ values shown, the spin gap does not open at half filling 
in Figs.~\ref{fig:HH_phase_diagram} (a) and (b).
Interestingly, a spin gap is induced by doping in these two cases.
The phonon-assisted backward scattering [$Y_1(l)$] is destructively 
interfered with the Umklapp process [$Y_\rho(l)$] so that it is 
not strong enough to open a spin gap at half filling for the parameters 
used in the figures.
Upon doping, the destructive interference becomes weak and 
the spin gap opens.
Upon further doping, the effects of 
$Y_\rho(l)$ and $Y_1(l)$ are overcome by the phonon-assisted forward scattering 
[$Y_2(l)$] in Eq.~(\ref{eq:x_rho_l}) 
so that $X_\rho^\ast > 0$, $K_\rho^\ast > 1$, and therefore the superconductor 
correlation becomes dominant.
As the phonon frequency $\omega$ increases, the SG-CDW phase disappears 
from the 
higher doping side and finally goes away for a 
phonon frequency at a fraction of the Fermi energy.
The fragility of the SG-CDW phase in the Holstein-Hubbard model is due to 
the fact that the cut off of the Umklapp process acts both for 
opening a spin gap and for strengthening the superconductor correlation.
Because of retardation, the former effect becomes largest at energy 
of order $\omega$, while the latter effect is independent of retardation.
This is why the SG-CDW phase is possible only for small $\omega$.
The filling dependence becomes weak with increasing $\omega$.

The above arguments become clear when one observes the evolution
of renormalized quantities with decreasing energy scale
shown in Fig.~\ref{fig:HH_evolution} for the Holstein-Hubbard model.
The first zero of $J_0(x)$ is about 2.4 so that the Umklapp process
is cut off at about $l \sim 2$ [Fig.~\ref{fig:HH_evolution} (c)].
Before reaching the cut-off energy scale, $X_\rho(l)$ rapidly decreases
[Fig.~\ref{fig:HH_evolution} (b)] and
$Y_1(l)$ also decreases [Fig.~\ref{fig:HH_evolution} (d)]
due to the destructive interference with
$Y_\rho(l)$, which is not shown but grows exponentially.
Since the phonon frequency is very small here, the electron-phonon interaction
becomes effective at a much smaller energy scale.
The products of an electron-phonon coupling and the corresponding phonon
propagator have a peak around $l \sim 7$ and decrease exponentially after
the peak [Figs.~\ref{fig:HH_evolution} (e) and (f)].
Before reaching $l \sim 7$, $X_\sigma(l)$
decreases proportionally to the inverse of $l$ but is still positive
[Fig.~\ref{fig:HH_evolution} (a)].
The magnitude of the slope of $X_\sigma(l)$ becomes large again
around $l \sim 7$ and finally $X_\sigma(l)$ becomes negative:
it flows to $-\infty$.
Because of the destructive interference, $Y_1(l)$ is smaller than $Y_1$
around $l \sim 7$, and consequently $Y_1(l) D_\pi(l)$ is smaller
than $Y_2(l) D_0(l)$.
Therefore, $X_\rho(l)$ increases after the Umklapp process is cut off.
However, the electron-phonon effect is not strong enough to make
$X_\rho(l)$ positive.

Phase diagrams for the SSH-Hubbard model supplemented with a forward scattering 
are shown in Fig.~\ref{fig:SHfs_phase_diagram}.
Now $Y_\rho(l)$ and $Y_1(l)$ are constructively interfered with each other.
The tendency for opening a spin gap near half filling is much stronger than 
in the Holstein-Hubbard model, 
regardless of supplementation with a forward scattering.
Note the difference between the scales of 
Figs.~\ref{fig:HH_phase_diagram} and 
\ref{fig:SHfs_phase_diagram}.
Contrary to the Holstein-Hubbard model, as the filling deviates from the half, 
the constructive interference becomes weak as well as the tendency for opening 
a spin gap.
The critical coupling strength for a spin gap is therefore an increasing 
function of $\mid \delta n \mid$, whose slope becomes gentle 
with increasing $\omega$.
Since $Y_\rho(l)$ and $Y_1(l)$ are enhanced by each other, 
larger deviation from half filling is necessary for $Y_2(l)$ to overcome 
these two to make the superconductor correlation dominant, compared with 
the Holstein-Hubbard model.
As a consequence, the SG-CDW phase appears in a much wider region.
As the phonon frequency $\omega$ increases, 
the interference effect becomes weak and larger electron-phonon coupling 
is necessary for a spin gap.
Therefore, the SG-CDW phase is shifted to larger $Y_{\rm ph}$ values 
with increasing $\omega$.
However, this phase remains in the $\omega \rightarrow \infty$ limit.
In this limit, the condition for a spin gap is the same as in 
the Holstein-Hubbard model, while the condition for a dominant 
superconductor correlation is different due to the sign factor $f$ 
in the initial condition for $Y_\rho(0)$.

The evolution of renormalized quantities with decreasing energy scale
is shown in Fig.~\ref{fig:SHfs_evolution} for the SSH-Hubbard model
supplemented with a forward scattering.
Before the Umklapp process is cut off at about $l \sim 2$
[Fig.~\ref{fig:SHfs_evolution} (c)], $X_\rho(l)$ rapidly decreases
[Fig.~\ref{fig:SHfs_evolution} (b)].
Note that $Y_{\rm ph}$ is set at 0.1 in Fig.~\ref{fig:SHfs_evolution},
while it was set at a much larger value of 0.4 in Fig.~\ref{fig:HH_evolution}.
Because of the constructive interference with $Y_\rho(l)$,
$Y_1(l)$ increases when the Umklapp process is effective
[Fig.~\ref{fig:SHfs_evolution} (d)].
Before the electron-phonon effect becomes largest at about $l \sim 5$
[Figs.~\ref{fig:SHfs_evolution} (e) and (f)],
$X_\sigma(l)$ decreases roughly proportionally to the inverse of $l$
[Fig.~\ref{fig:SHfs_evolution} (a)].
The magnitude of the slope of $X_\sigma(l)$ becomes large again
around $l \sim 5$ and finally $X_\sigma(l)$ becomes negative and flows to
$-\infty$.
Compared with Fig.~\ref{fig:HH_evolution}, the products of
an electron-phonon coupling and the corresponding phonon propagator
grow more rapidly and are not negligible when the Umklapp process
is cut off.
Therefore, the region where the electronic backward scattering
decreases $X_\sigma(l)$ and the region where the phonon-assisted
backward scattering decreases $X_\sigma(l)$ are not clearly
distinguishable.
Because of the constructive interference, $Y_1(l)$ is about twice
as large as $Y_1$ around $l \sim 5$, and consequently $Y_1(l) D_\pi(l)$
is about twice as large as $Y_2(l) D_0(l)$.
Therefore, the effects of the back and forward scatterings assisted by
phonons cancel each other [Eq.~\ref{eq:x_rho_l}]
so that $X_\rho(l)$ does not change so much
after the Umklapp process is cut off.

\section{SUMMARY}\label{sec:summary}

We have studied asymptotic correlation functions of nearly-half-filled 
one-dimensional conductors coupled with dispersionless phonons 
using the boson representation of the electron fields and 
the lowest-order renormalization group method.
On-site repulsion is considered for the electronic scattering processes.
As the renormalization proceeds, the Umklapp process is initially effective 
even away from half filling and this causes the filling dependence of 
the correlation functions.
We consider a phonon-assisted forward scattering which enables the 
superconductor correlation to be dominant. 
In addition, we take two types of phonon-assisted backward scatterings 
originating from the Holstein model and from the SSH model.
They couple with both spin and charge until the energy scale reaches 
the phonon frequency: variation of filling modifies the spin 
correlation also.
The variety of phases comes from 
the cut-off energy scale of the Umklapp process relative 
to that of electron-phonon interaction, which depends upon doping.
The variety also comes from 
the interference of the Umklapp process and the phonon-assisted backward 
scattering, which is destructive for the Holstein-type on-site 
electron-phonon coupling and constructive for the SSH-type intersite 
electron-phonon coupling.

At half filling, a charge gap opens by the Umklapp process in most cases.
A spin gap also opens if the phonon-assisted backward scattering is 
strong enough.
Since these two processes interfere constructively for the SSH coupling, 
the critical coupling strength for a charge density wave 
with both spin and charge gaps is weaker for the SSH-Hubbard model than for 
the Holstein-Hubbard model. 
The difference between these two models is large for small phonon frequency 
since the interference is effective for a wider energy range of 
the renormalization process.
This interference effect appears also near half filling, where 
the Umklapp process is integrated out at finite energy so that 
the charge excitation spectrum is always gapless.
The spin gap opens more easily by the intersite coupling than by the 
on-site coupling, especially for small phonon frequency.
The above two processes enhance the density-wave correlation.
To make the superconductor correlation dominant, the phonon-assisted 
forward scattering has to overcome them.
It is achieved upon doping much easier for the Holstein-Hubbard model 
since the two negative processes interfere destructively.

Far away from half filling, a phase diagram is divided basically into 
the gapless metallic phase with a dominant density-wave correlation, 
if the on-site repulsion is stronger, and 
the conventional superconductor phase with a spin gap, 
if the electron-phonon interaction is stronger. 
Near half filling, we found a spin-gap metallic phase with a dominant 
charge-density-wave correlation between the two phases.
This phase appears in a wider range for the SSH-Hubbard model supplemented 
with a forward scattering because of the constructive interference 
as mentioned above.
For the Holstein-Hubbard model, it appears only for small phonon frequency, 
where the spin gap is induced by doping since the destructive interference 
is weakened as the filling deviates from the half.

For the spin-gap metallic phase, one-dimensionality is essential since 
the spin excitation and the charge excitation are decoupled in the 
asymptotic limit. 
The $g_{1\perp}$ process is effective only for the spin excitation and 
produces a gap if strong.
For higher dimensions, a considerably good nesting would be necessary 
for a similar phase.

\acknowledgments
This work was supported by Grant-in-Aid for Scientific Research 
on Priority Areas 
``Anomalous Metallic State near the Mott Transition'' and 
``Novel Electronic States in Molecular Conductors'' 
from the Ministry of Education, Science, Sports and Culture.

\begin{figure} \caption{
Phase diagrams of 
(a) the Holstein-Hubbard model and
(b) the SSH-Hubbard model supplemented with a forward scattering, 
at half filling $\delta n = 0$ and without magnetization $m = 0$.
The electronic parameter is set at $Y_{\rm el} = 0.4$.
}\label{fig:half_filling}
\end{figure}

\begin{figure} \caption{
Phase diagrams of the Holstein-Hubbard model for 
(a) $\omega = 10^{-3} E_{\rm F}$, 
(b) $\omega = 10^{-2} E_{\rm F}$,
(c) $\omega = 10^{-1} E_{\rm F}$, and 
(d) $\omega = \infty$, 
for $\delta n \neq 0$ 
(See Fig.~1 for $\delta n = 0$.)
and $m = 0$.
The electronic parameter is set at $Y_{\rm el} = 0.4$.
}\label{fig:HH_phase_diagram}
\end{figure}

\begin{figure} \caption{
Phase diagrams of the SSH-Hubbard model 
supplemented with a forward scattering for 
(a) $\omega = 10^{-2} E_{\rm F}$, 
(b) $\omega = 10^{-1} E_{\rm F}$, 
(c) $\omega =         E_{\rm F}$, and 
(d) $\omega = \infty$,
for $\delta n \neq 0$ 
(See Fig.~1 for $\delta n = 0$.)
and $m = 0$.
The electronic parameter is set at $Y_{\rm el} = 0.4$.
}\label{fig:SHfs_phase_diagram}
\end{figure}

\begin{figure} \caption{
(a) $X_\sigma (l)$, (b) $X_\rho (l)$, (c) $2\pi\delta n (l) \alpha$, 
(d) $Y_1 (l)$, (e) $Y_1 (l) D_\pi (l)$, and (f) $Y_2 (l) D_0 (l)$, 
in the SG-CDW phase of the Holstein-Hubbard model.
Parameters are 
$\omega = 10^{-3} E_{\rm F}$, 
$Y_{\rm ph} = 0.4$, 
$2\pi\delta n \alpha = 0.1$, 
$m = 0$, and 
$Y_{\rm el} = 0.4$. 
}\label{fig:HH_evolution}
\end{figure}

\begin{figure} \caption{
(a) $X_\sigma (l)$, (b) $X_\rho (l)$, (c) $2\pi\delta n (l) \alpha$,
(d) $Y_1 (l)$, (e) $Y_1 (l) D_\pi (l)$, and (f) $Y_2 (l) D_0 (l)$,
in the SG-CDW phase of the SSH-Hubbard model 
supplemented with a forward scattering. 
Parameters are
$\omega = 10^{-2} E_{\rm F}$,
$Y_{\rm ph} = 0.1$,
$2\pi\delta n \alpha = 0.25$,
$m = 0$, and 
$Y_{\rm el} = 0.4$.
}\label{fig:SHfs_evolution}
\end{figure}

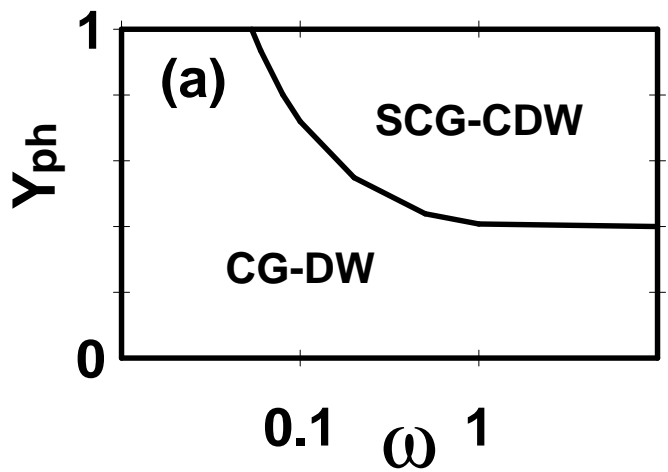
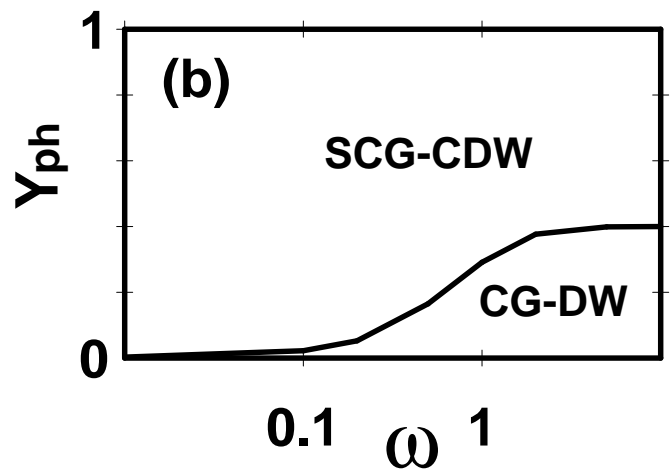
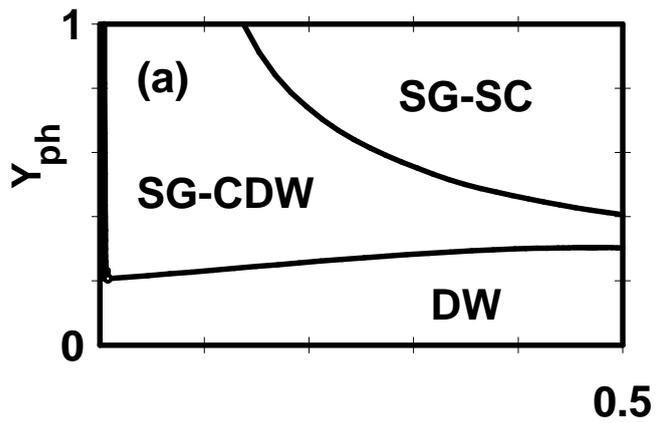
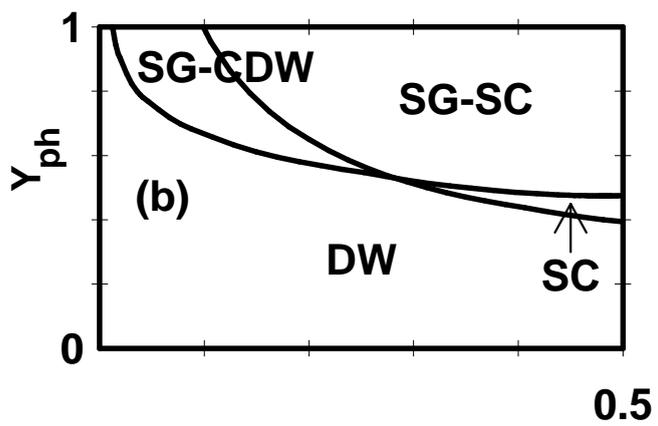
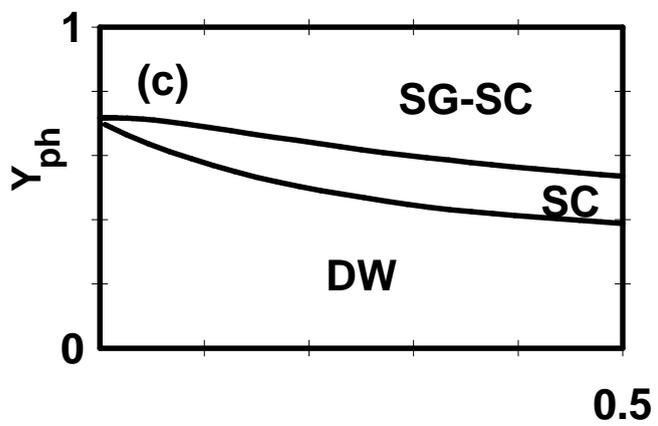
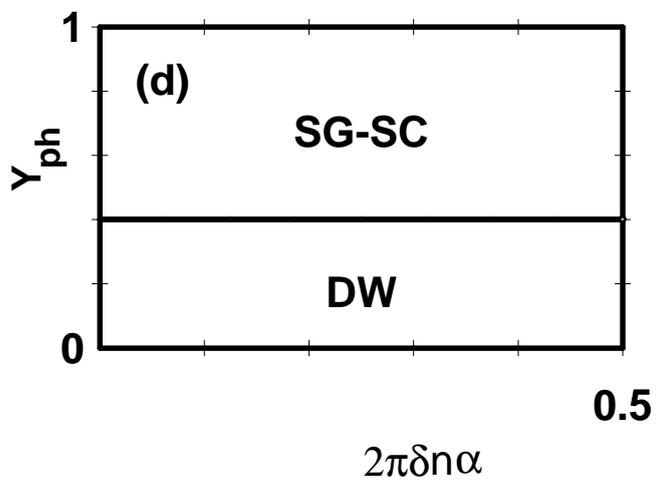
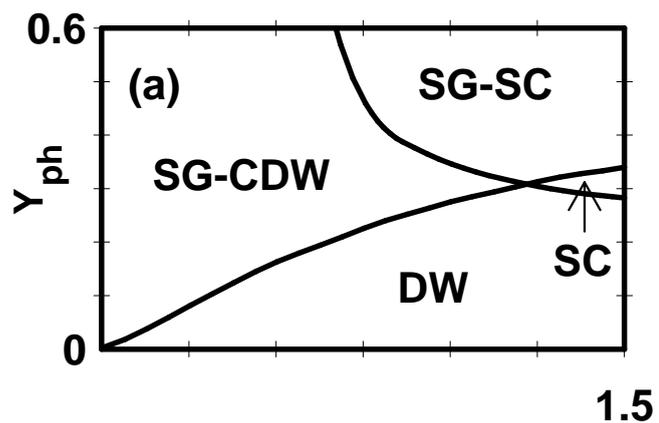
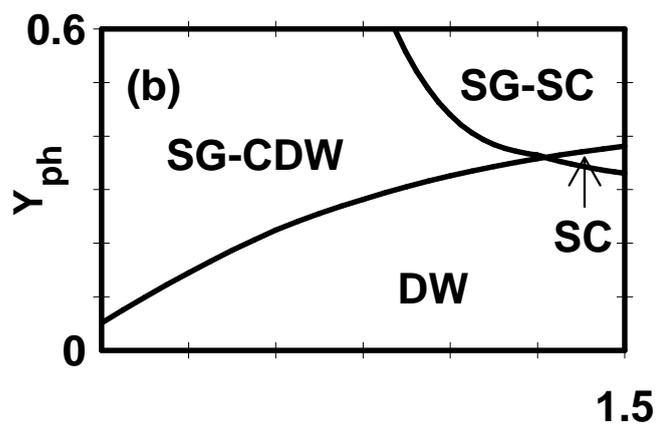
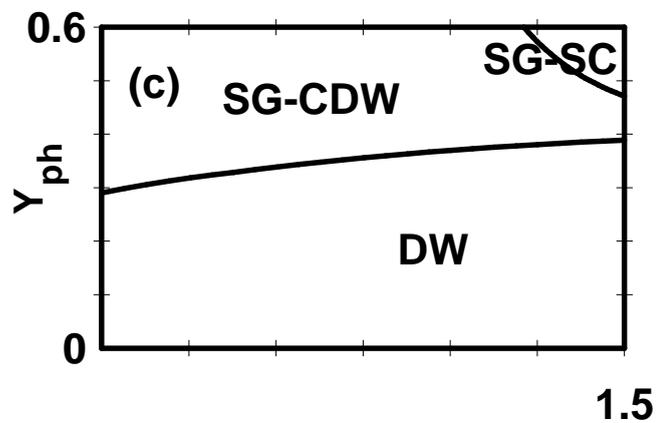
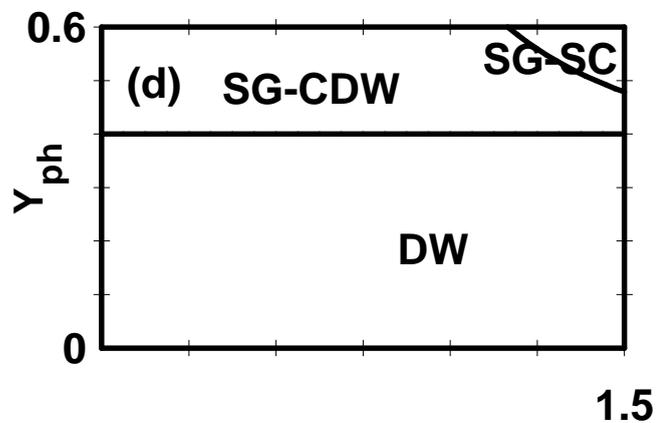
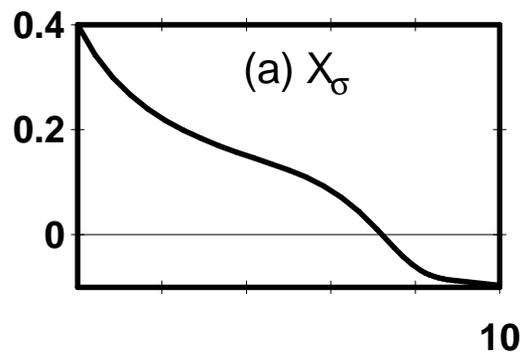
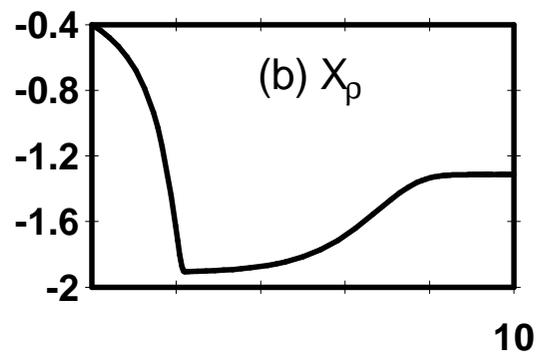
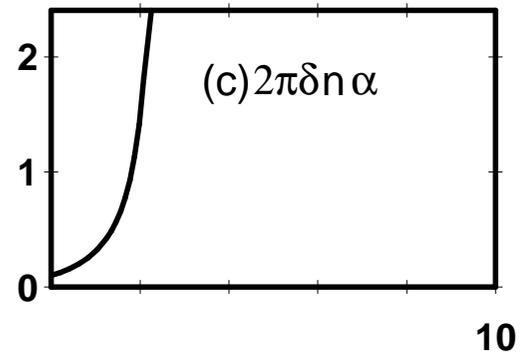
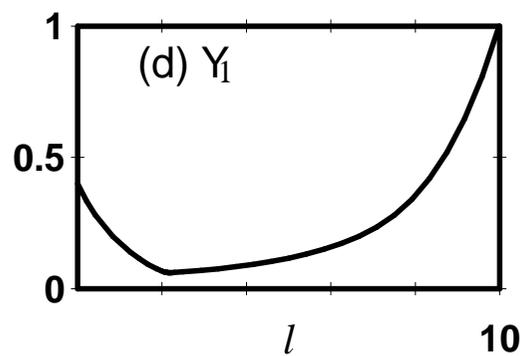
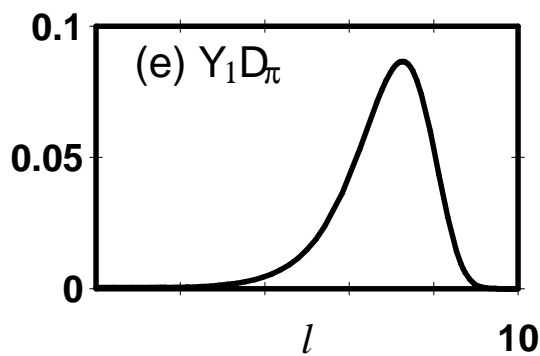
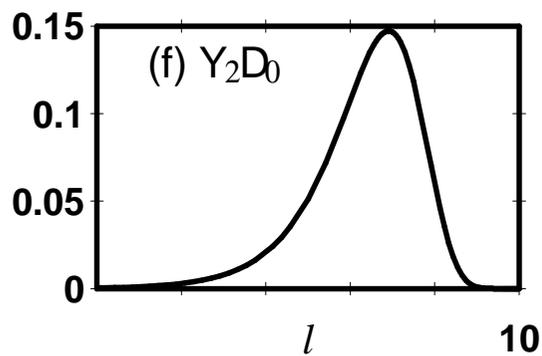
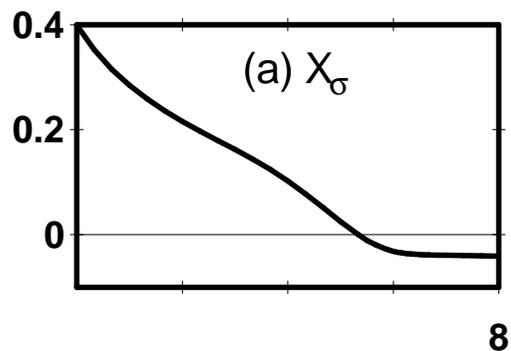
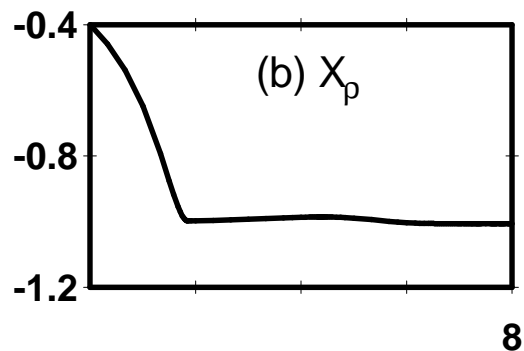
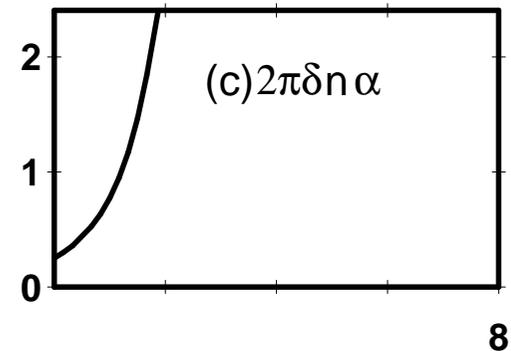
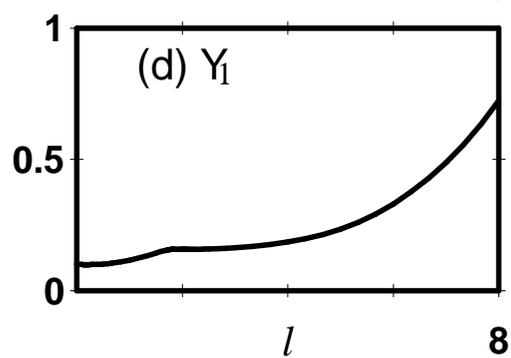
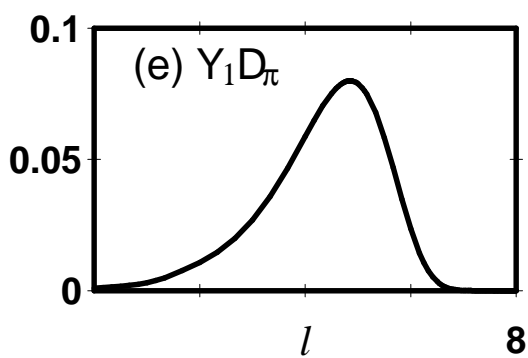
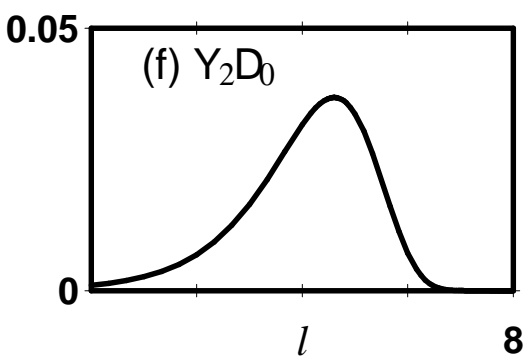
\end{document}